\documentstyle[12pt]{article}
\newcommand{\ageq}{\mbox{\
\raisebox{-.9ex}{$\stackrel{\textstyle >}{\sim}$}\ }}

\def\eps{{\epsilon}}
\def\begineq{\begin{equation}}
\def\endeq{\end{equation}}

\newcounter{saveeqn}

\newcommand{\alpheqn}%
{\setcounter{saveeqn}{\value{equation}}\setcounter{equation}{0}%
\renewcommand{\theequation}%
{\mbox{\arabic{saveeqn}\alph{equation}}}%
\addtocounter{saveeqn}{1}}
\newcommand{\reseteqn}{\setcounter{equation}{\value{saveeqn}}%
\renewcommand{\theequation}{\arabic{equation}}}
\renewcommand{\theequation}{\arabic{equation}}

\begin{document}
\bibliographystyle{prsty}
\title{Scaling and Linear Response in the GOY Turbulence Model}

\author{Leo Kadanoff $^{1}$, Detlef Lohse $^{1,2,*}$,
and Norbert Sch\"orghofer $^{1}$}
\date{\today}
\bigskip

\maketitle

\bigskip

\centerline{$^{1}$ The James Franck Institute, The University of Chicago,}
\centerline{ 5640 South Ellis Avenue, Chicago, IL 60637, USA}

\centerline{$^{2}$ Fachbereich Physik, Universit\"at Marburg,}
\centerline{Renthof 6, 35032 Marburg, Germany}

\bigskip

%\date{}
\maketitle
\bigskip
\bigskip
\bigskip

\begin{abstract}

The GOY model is a model for turbulence in which two conserved quantities
cascade up and down a linear array of shells.  When the  viscosity parameter,
$\nu$,  is small the model has a qualitative behavior which is similar to the
Kolmogorov theories of turbulence. Here  a static solution to the model is
examined, and a linear stability analysis is performed to obtain response
eigenvalues and eigenfunctions.  Both the static behavior and the linear
response show an inertial range with a relatively simple scaling structure. Our
main results are: (i) The response frequencies cover a wide range of
scales, with
ratios which can be understood in terms of the frequency scaling properties of
the model. (ii)  Even small viscosities play a crucial role in determining the
model's eigenvalue spectrum. (iii) As a parameter within the model is varied, it
shows a ``phase transition'' in which there is an abrupt change in many
eigenvalues from stable to unstable values.   (iv) The abrupt change is
determined by the model's conservation laws and symmetries.

This work is thus
intended to add to our knowledge of the linear
response of a stiff dynamical systems and at the same time to help illuminate
scaling within a class of turbulence models.

\end{abstract}

\vspace{0.5cm}\noindent

\newpage

\section{Introduction}

In turbulent flow a hydrodynamic system couples together many different
length scales and thus shows in a single process a huge range of relaxation
rates. There are a variety of simplified models of turbulence which are also
intended to show this wide range of frequency and wave-number scales. One
such model goes under the inelegant acronym of "GOY". The model couples
together a large number of shells, each with its characteristic scale of
wave-vectors and relaxation times.  Shells are spaced logarithmically in wave
vector. The $n$th shell is characterized by a single complex velocity, $U_n$,
which then depends upon time, $t$. The model is a linked set of ordinary
differential equations for all these $U_n$ with equations picked to mimic those
of real hydrodynamic flow.

We do not know whether the model has much to do with turbulence. But it
certainly illustrates the behavior of a stiff
system.  (Stiff systems are ones in which numerical simulations are made
difficult by the effects of a huge range of relaxation rates.)  In this
paper, we describe the time dependence of the model in the simplest possible
situation, the linear response to disturbances around a static solution. The
response is described as an eigenvalue problem, with the response matrix being
a large matrix which inherits the conservation laws, the scaling properties, and
the symmetry principles of the GOY model. We look for the eigenvalues and
eigenstates of the matrix. As we find them we see that the linear response, in
turn,  shows a considerable richness including scaling
behavior and the analog of a phase transition. Much of this behavior can be
understood in terms of the several symmetries and conservation
laws of the original model.

The richness of behavior in this linear response theory serves to remind us
that there is
one area of scaling or similarity theory which has not been
fully explored, the determination of eigenfunctions for matrices which have an
underlying scaling structure.  We also do not understand very much about
turbulence, or even about the flow of information up and down dynamical
linear chains.  This paper is about these three not-fully-understood
areas of applied mathematical science.

To start out, we show the most interesting results of our study.  We plot in
figure \ref{spec}, the eigenvalues of the linear stability in a sort of polar
diagram in which the polar coordinates,  $\theta$ and $r$ respectively are the
eigenvalues' phase and are
proportional to the logarithm of the magnitude of the eigenvalue. (See equation
(\ref{r}) below for a precise definition.){\footnote {Similar plots are given for
different values of $\eps$ in figures (\ref{spec.5}) and (\ref{specc}). Their
discussion will be postponed to later on in the paper.}}  The two different
parts of the plot show the response to a purely real disturbance (in part
\ref{spec}a) and to a purely imaginary disturbance in part \ref{spec}b.
This distinction is meaningful because both the basic equations and the static
solution are purely real. For this figure we have picked a particularly small
value of the viscosity parameter so as to arrive at a simple scaling behavior.
Indeed the simple spacing of the points in figure 1a shows that a simple
multiplicative law generates the higher order eigenvalues from the lower order
ones.  That is why the points fall on a straight line with regular spacings.
There is a tremendous range of scaling of the eigenvalues, and it all looks
provocatively simple. This paper is mostly aimed at producing a partial
explanation of these figures.

The paper starts with the introduction of the GOY model, and treats scaling
for the static solution in section 2. In section 3 we discuss the linear 
stability analysis
in general and apply it to the consideration of the scaling of the simpler,
real component of the response. Also a relation between eigenvalues and 
conservation laws is derived. Section 4 discusses the linear stability in the
imaginary response sector and the phase transition and scalings seen there.
Section 5 summarizes the results. Footnotes
and appendices discuss findings that are not necessary for the main line of
thought.

\subsection{The model}

The basic structure of GOY originated from Gledzer \cite{gle73}, and was
motivated by the cascade structure of turbulent eddies and conservation laws.
Later, Ohkitani and Yamada \cite{yam87}
generalized the model and carried out numerical studies
that revealed chaotic behavior and a dynamic
scaling of velocity fluctuations.
These studies have been extended in\cite{jen91,ben93,bif95,kad95}.
A popular introduction can be found in \cite{kad95b}.

The basic ingredient is the hierarchical structure: The $n^{th}$ shell is
characterized by a wave vector of length
\begineq
k_n=k_0 \lambda^n, \qquad n=1,2,\dots N,
\label{hierachy}
\endeq
with $\lambda > 1$ and by a complex velocity mode $U_n$. The Navier-Stokes
dynamics is mimicked by the following set of ODEs:
\begineq
{d\over dt}U_n=F_n  -  C_n(U^*) -D_n,
\label{eq1}
\endeq
where the three terms represent respectively forcing and cascade processes
and dissipation. (The
star $^*$ indicates complex conjugation.)

We pick a forcing on the first shell
\begineq F_n= \delta_{n,1} f. \endeq
Most previous studies
\cite{yam87,jen91,kad95,sch95} of the GOY model  use a forcing on the fourth
shell. Our choice of the first shell seems to give a simpler structure to the
results. More details can be found in Appendix A.  In our numerical work we
shall choose $\lambda=2, k_0=\lambda^{-1}, f=1$ unless otherwise stated.

The dissipation term is \begineq -D_n=-\nu k_n^2U_n
\endeq
and is the $k$-space representation of the usual viscous dissipation process.

The cascade term couples the shell $n$ to its nearest and  next
nearest neighbors,
\begineq
C_n(U)=
k_{n}U_{n+1}U_{n+2}-\eps k_{n-1}U_{n-1}U_{n+1}-(1-\eps)k_{n-2} U_{n-1}U_{n-2}.
\label{eq2}
\endeq
The boundary conditions are simply $U_{-1}=U_0=0$ and that the
velocities go to zero as $n$ becomes large. (In our numerics
we represent that by cutting off at some large shell numbered
$N$ and then using the conditions $U_{N+1}=U_{N+2}=0$.)  Because the
static equations couple $n$ with the four immediately neighboring shells,
we need
four boundary conditions to define the problem.
Two of the conditions are at the low-$n$ end, and two at the high. The nature
of the boundary conditions will become crucial later on.

The model parameter $\eps$ determines the ratio between upscale and downscale
coupling. It gives us a convenient tool for varying the model and seeing
qualitatively different ranges of behavior.

In the inviscid, unforced limit there are two conserved quantities: the energy
\alpheqn
\begineq
E={1 \over 2} \sum_n{|U_n|^2}
\label{cons1}
\endeq
and a second conserved quantity \cite{kad95} of the form
\begineq
H=\sum_n{|U_n|^2 \over (\eps-1)^n}
\label{cons2}
\endeq
\reseteqn
that can be roughly associated with the helicity in fluid motion \cite{kad95}.
Even though the association is not perfect, we shall call this quantity the
helicity in this paper \cite{ben96}.

The GOY model shows dynamical as well as static behavior.
There is a static solution \cite{sch95} of the GOY model
($dU/dt=0$) in which the phases are zero. When we wish to point particularly
to the static solution, we shall write it as $u_n$, while we use $U$ to refer
to either the static or the dynamic case. In this paper, we study the
static solution, $u_n$,  and its linear stability properties.

The phases are not uniquely fixed for $u_n$ \cite{sch95}
which is part of a one parameter family of solutions. (We pick the particular
solution which has $u_n$ real and positive.) The invariance in the phase will
be important in our linear stability analysis. It gives rise to a zero
eigenvalue.  There are other approximate invariance properties which will be
discussed when we get to the linear stability analysis.

\section{Scaling (of static solution)}
We are interested in understanding the behavior of the model in the limit
as the viscosity becomes small.  In that limit, there are three regions of
$k$-space, or of $n$, called the
stirring subrange SSR, the inertial subrange ISR, and the viscous subrange VSR.
The last is dominated by the viscosity term and has a velocity which decays
very rapidly with $k$. (In the GOY model, the decay is exponential in a
power \cite{sch95} of $k_n$.) The SSR is naturally enough the range
 in which stirring is
directly important. In our case this comprises only $n=1$.   The inertial subrange
is the intermediate range of wave vectors between these two. Here, the behavior
is dominated by the cascade term. In the center of the ISR,  the solution
is best described by using the product of three velocities:
\begineq  \Delta_n(u)=k_n U_n
U_{n+1} U_{n+2}.
\label{del}
 \endeq
In the region in which only the cascade term matters, this product is
 \begineq
\Delta_n=A+B (\eps-1)^n
\label{delta}
\endeq
where A and B are adjustable constants of integration, representing
respectively the energy and helicity flux though the inertial range. We shall
work with $\eps$ in the unit interval so that, for large $n$, the A term
dominates. In this region: \begineq
U_n=W_n k_n^{-1/3}
\label{2.2}
\endeq
where $W_n$ is a periodic function with period three. This behavior
continues as $n$ increases until one enters the dissipative range, and
$U_n$ begins to fall off very rapidly.

Notice that we have, as expected, four undetermined parameters in the solution.
In the ISR these parameters are $A$ and $B$ and the two parameters which define
the period-three oscillations.

The scaling behavior is illustrated in figure \ref{sol} which plots
$W_n=k_n^{1/3} U_n$
against $n$ for a case in which $\nu$ has the value $10^{-3}16^{-20}$.
Notice how $U_n$ shows a simple scaling for large $n$ up to $n$ of about 67, and
then it nose-dives.
The nose-dive occurs as $n$ increases above a dissipative
threshold, which is achieved when the dissipation term and the cascade terms are
roughly of equal size. Usually, $W_n$ is of the order unity in the inertial range
so that this dissipative cutoff can be computed as a function of a threshold
value of wave vector, called $k_D$, which obeys
\begineq
k_D^{-4/3} \sim \nu .
\label{2.3}
\endeq
The condition is thus that the viscous effects should dominate for values
of $n$ larger than this dissipative threshold, that is,
\begineq
n>N_D \sim -{3\over 4} log_{\lambda} \nu
\label{2.4}
\endeq

There is a simple scaling theory which we can apply to this case.
Because the static solution has an asymptotic period three \cite{kad95}, the
system should be almost unchanged by the transformation {\footnote{ The change
in $N$ would be unimportant were we really working with very large values of
$N$.  For numerical convenience we work with N only seven or eight larger than
$N_D$.  This change in $N$ in equation (\ref{2.5a}) eliminates effects
produced by
changing the cutoff. Throughout this paper we use a prime to denote
quantities changed by the transformation of equations (\ref{2.5a}) and
(\ref{2.5b}).}:
\alpheqn \begineq
\nu \to \nu'=\nu/\lambda^4
\label{2.5a}
\endeq
\begineq
N \to N'=N+3
\label{2.5b}
\endeq
\reseteqn
In particular, as one goes from the primed to the unprimed
situation, the static solution should be basically the same at both ends, with
the large n-end being only modified by having smaller values of $u$. Let $u'$ be
the solution for the velocity in the situation changed according to equations
(\ref{2.5a},\ref{2.5b}).  The two scaling symmetries can be written as:
\alpheqn \begineq \hbox{(S)}\quad u'_n=u_n(1+\it{O}(\frac{k_n}{k_{N_D}})^{2/3})
\quad\hbox {for}\quad N_D-n>>1, \label{2.6a} \endeq
\begineq
\hbox{(L)}\quad \lambda u'_{n+3}=u_n(1+\it{O}(\eps-1)^{n})
\quad\hbox{for}\quad n>>1.
\label{2.6b} \endeq
\reseteqn
In these equations ``S'' stands for small $k$  and ``L'' refers to large
$k$. Note that the first of these equations remains valid in the SSR while the
latter remains valid in the VSR.

In equations (\ref{2.6a},\ref{2.6b}) we have included estimates of the error
resulting from the terms we have not taken into account in constructing the
scaling.  In the small $k$-range, we do not include viscous effects, so the
error estimate is the relative size of the viscous term.  For large $k$,
we do not include the helicity flux represented by the $B$ in equation
(\ref{delta}) so this effect is put into the error.  The global error is set by
the sizes of each of these errors at the far ends of the viscous subrange. One
such error is the value of $\nu$.  The other, and usually larger effect, is the
effect of the helicity flux term, $B$, at the large-$n$ end of the ISR.
This term has the order of magnitude,
\begineq error = \it{O} ( 1-\eps )^{N_D}.
\label{err}
\endeq
This scaling error must be at least as large as the maximum of this error
and $\nu$.

In terms of $W$, defined by equation (\ref{2.2}) the two scaling equations
imply respectively that $W'_n=W_n$ and that $W'_n=W_{n+3}$. In the center of the
ISR both of these scaling symmetries are valid. Then $W$ has a period three
symmetry:
 \begineq
\hbox{(I)} \quad W'_n=W_n=W_{n+3}
\quad\hbox{for}\quad n>>1 \quad\hbox{and}\quad N_D-n>>1
\label{intermed}
\endeq
``I'' refers to an intermediate range which occurs in the middle of the ISR.

To check our thinking, we calculate the static solution of the GOY model.
Deviations from
scaling can be seen by looking at the behavior of
\alpheqn
\begineq
\delta_{S,n} = 1-{W_n \over W_n'},
\label{delus}
\endeq
\begineq
\delta_{L,n} = |1-{W_n \over W_{n+3}'}|.
\label{delul}
\endeq
\reseteqn
Both quantities are plotted in
figure \ref{solerr}.
Also shown in this figure are theoretical lines which show the errors
defined in equations (\ref{2.6a}),(\ref{2.6b})  and (\ref{err}). Theory and
experiment show excellent agreement.

\section{Eigenvalue spectrum}
\subsection{Linear Response}
The next stage is to do linear stability analysis. We  consider small
deviations about the static solution $u_n$ by writing:
\begineq
U_n(t) = u_n (1+ \delta \Phi_n e^{\sigma t})
\label{dev}
\endeq
Since the static solution $u$ is real
the eigensolutions split neatly into oscillations of the phase and  the
amplitude, corresponding respectively to $\delta \Phi_n$ being real and
imaginary.
Then we have an eigenvalue equation
 \begineq
\sigma \delta \Phi_n = \sum_{m} A_{nm} \delta \Phi_m.
\label{eq3a}
\endeq
We distinguish the two different cases with subscripts $M$ for magnitude and
$\phi$ for phase. We use a superscript $j$ to denote which eigenvalue is being
considered.   In
what follows we will always order the eigenvalues so that the magnitude of
the eigenvalue increases with increasing values of the index $j$.
Then the eigenstate will be very small for those shells which have $n$
considerably smaller than $j$. We will then argue that the behavior of the
response matrix for $n$ and $m$ of order $j$ will play a large role in
determining the eigenvalue.

Note that the response matrix $A$ has two parts. The
dissipative part is
\alpheqn
\begineq
D_{nm} =\nu \delta_{nm} k_n^2,
\label{eq5a}
\endeq
and the cascade response is:
\begineq
C_{nm} = u_m {\partial \over \partial u_m} C_n (u).
\label{eq5b}
\endeq
\reseteqn
Now the structure of $A$ is different for the two kinds of
response. For the magnitude response,
\alpheqn
\begineq
A_M = -C -D,
\label{eq4a}
\endeq
while for the phase response
\begineq
A_\phi =  C -D.
\label{eq4b}
\endeq
\reseteqn

{}From (\ref{eq2}) one sees that the matrix $-C$ has rows of the form
\begineq
0,\ldots,c k_n u_{n-1}u_{n-2},k_n(b u_{n+1}+c u_{n-2})u_{n-1},0,
k_n(u_{n+2}+b u_{n-1})u_{n+1},k_n u_{n+1}u_{n+2},0,\ldots,
\label{matrix0}
\endeq
where $b=-\eps/\lambda$ and $c=(\eps-1)/\lambda^2$.

\subsection{Eigenvalue spectra}
At the center of the ISR
$u_n$ shows a simple scaling superposed on top of a period three behavior.
Thus, the response matrix also has a period three scaling:
\begineq
C_{n+3,m+3}=\lambda^{2} C_{n,m} \quad\hbox{for $ n,m $ in region I}.
\label{eqwrong}
\endeq
Consequently, if $C$ dominates the behavior of the matrices $A$ in the
determination of eigenvalues in some range of $j$, then the eigenvalues would
obey the scaling property
\alpheqn
\begineq
\sigma^{j+3}_M=\lambda^{2} \sigma^{j}_M  \quad\hbox{for $j$ in region I}
\endeq
\begineq
\sigma^{j+3}_\phi=\lambda^{2} \sigma^{j}_\phi \quad\hbox{for $j$ in region I}
\label{sigscale}
\endeq
\reseteqn
In region I, then,  the logarithms of the eigenvalues should be evenly
spaced along straight lines with spacing of $\log \lambda^2 $. Also, we might
think that if  $C$ dominates the  behavior of the matrices $A$, then the phase
and magnitude eigenvalues should be the same except for a minus sign. (See
equations (\ref{eq4a}) and (\ref{eq4b})).
The viscosity is unimportant in the entire
region S.  Thus we expect  \begineq \sigma^{j}_M= - \sigma^{j}_\phi
\quad\hbox{for $j$}\quad  \hbox{in region S}. \label{badeq}
\endeq

Figure \ref{spec} shows the eigenvalue spectra for $\eps=0.3$
for both modulus and
phase  stability matrices $A_M$ and $A_\phi$.
As the distance between the eigenvalues
$\sigma^{j}$ in the hierarchical GOY model
grows roughly exponentially, it is hard to visualize the spectra in
the complex plane.
To have some kind of visualization we use a kind of polar representation in
which the phase of the plotted point is exactly the phase of $\sigma$ while
the distance from the center of the polar plot is given by: \begineq
r =\log_\lambda(1+2^{10} |\sigma|).
\label{r}
\endeq
The factor of $2^{10}$ is put in to enhance the visibility of eigenvalues with
small values of $|\sigma |$. The plot has unstable eigenvalues showing up on
the right hand side of the origin while stable ones show up on the left. For
large eigenvalues, even spacings on the plot mean that successive
eigenvalues have
ratios which are a constant.  Thus even spacings are indicative of some kind of
simple scaling.

In both the magnitude and the phase sector, the
eigenvalues are arranged in several branches. Within each branch there are
regions of even spacing, indicative of simple scaling. The magnitude 
eigenvalues seem particularly simple with three well-defined
branches: A set of 
real eigenvalues and a pair of complex conjugate branches. All eigenvalues are
stable. The phase eigenvalues show a more complex structure with what looks
like more regions of simple scaling.  Nonetheless both sets of eigenfunctions
show the even spacing demanded by scaling.  
{\footnote{
The existence of three branches is not due to the existence of a period
three in the solution. One can assume an approximate solution $u_n=k_n^{-1/3}$,
but the corresponding matrix also consists of three radial
branches, although we have no period 3 in the "solution" any more.
}}
As one might expect, the minimal
modulus eigenvalue  is  about of the size of the entries in the first row of the
cascade matrix $C$, the modulus of the second one is of the order of the second
row, and so on. Generally, the modulus of the nth eigenvalue is about of the
size of the nth row of the matrix. This rule holds roughly until the eigenvalue
with $j$ equal to $N_D$ is reached whereupon the successive eigenvalues are real
and have values of the order of the diagonal elements in the dissipation
matrix, $D$. In the center of the inertial range 
the eigenvalues change from real to a complex
conjugated pair and back to real and so on with a period three.

Thus, much of what we see is what we might expect. But not all. Equation
(\ref{badeq}) is completely inconsistent with the pictures we are seeing. This
equation implies that if we have stable eigenvalues in the magnitude sector,
we should have unstable ones in the phase sector.  But all eigenvalues in
figure
\ref{spec} are stable.  This result cannot be consistent with the notion that
viscosity is unimportant for the eigenvalues in some region of the response. We
have other difficulties.  The pattern of phase response eigenvalues looks much
more complex than the pattern of magnitude eigenvalues. Why should that be so?

Another difficulty is associated with the
prediction of Biferale \cite{bif95b}\cite{bif95c} that there would be a change
in behavior at  the special value of $\eps$  for which the contributions to the
helicity sum in equation (\ref{cons2}) grow toward the high-$n$ end of the
inertial range. The value is \begineq
 \eps_{bif}:= 1-\lambda^{-2/3}.
\label{luca}
\endeq
We have just seen that the model is stable for $\eps=0.3$, which lies below the
Biferale value $\eps$ approximately $0.37$. Look at figure \ref{spec.5}, which
is the analog of Figure \ref{spec} but now for $\eps=0.5$. The magnitude
spectrum
looks much the same as before, but there is a qualitative change in the phase
spectrum. Now, there are a large number of unstable eigenvalues in the phase
spectrum.  We interpret what we are seeing by saying that there is a qualitative
change in properties of the asymptotic ($\nu$ goes to zero) model at
$\eps=\eps_{bif}$. For $0<\eps<\epsilon_{bif}$ there are at most a finite
number of unstable eigenvalues.  At $\eps$ just above  $\eps_{bif}$ the system
acquires an infinite number of unstable eigenvalues. Figure \ref{specc} shows
the behavior of the spectrum at the Bifarale point.  Once again the phase
spectrum has a qualitatively new character, while the magnitude spectrum
remains much the same as before. We shall want to understand better how this
behavior arises from the linear stability analysis.

\subsection{Eigenvalues in the Magnitude sector}
The scaling behavior of the eigenvalues in the magnitude sector is given by
the very same ideas which we already used in our analysis of $u_n$. As we
shall see, some new ideas will be required for the phase sector. For this
reason, we shall dispose of the magnitudes here and move on to a more
extended discussion of phase eigenvalues in the next chapter.

The general structure of the eigenstates is illustrated in
figure \ref{someevs}.  Here we look at right eigenvectors with $j=26$. The
eigenvalue equation in the magnitude sector is
 \begineq
\sigma^j \delta \Phi_n^j = -\sum_{m} (D_{nm}+ C_{nm}) \delta \Phi_m^j .
\label{eigenM}
\endeq
For small $n$ the eigenvector is very small. In fact, the eigenvector must
decrease rapidly as $n$ decreases to enable the eigenvalue term from swamping
the right hand side of the equation.  This decrease is a falloff from a plateau
which occurs  at $n$ about equal to $j$. At this point, the eigenvalue term and
the cascade term in equation (\ref{eigenM}) are about equal in size.  As $n$
increases still further, the three parts of the cascade term each become much
larger than the eigenvalue term. To ensure the cancellation of the different
parts of the cascade term the eigenvector settles down to an oscillation with
period three, which holds throughout the entire ISR.  Finally as $n$ enters the
VSR, the eigenstate once again falls off quite rapidly.
The behavior of the phase eigenstate is much the same, except that there is no
falloff in the VSR.

One can describe the same process in physical terms by saying that the
eigenvalue term adds energy to the system which then
cascades toward the VSR, where the energy is dissipated.

Now one can see why the dissipation plays such a large role in determining
the behavior of the eigenvalues. The deviation $\delta \Phi_n^j $ produces a
change in the conserved quantities, both the helicity and the energy.
This change cascades down through the ISR toward the VSR.
Because of the conservation of energy this cascade is not damped and
remains constant in size until the VSR is reached. Since the eigenvector
remains constant into the VSR, then the VSR behavior serves as a kind of
large $n$ boundary condition on the eigenstate.

A simple count shows how this works. For large $n$ in the ISR, $\delta \Phi_n
$ is periodic with period three.  Then, there are two parameters which
describe the wave function:
\alpheqn
\begineq
p_1^j(k) =  {\delta \Phi_{3k+1} \over \delta \Phi_{3k}}
\label{para1}
\endeq
\begineq
p_2^j(k) =  {\delta \Phi_{3k+2} \over \delta \Phi_{3k}}.
\label{para2}
\endeq
\reseteqn
In our previous work \cite{sch95} we found that there were two conditions
upon the large-$n$ velocities, required to keep these velocities from blowing
up deep into the viscous subrange. Two parameters, two conditions. Everything
is determined.  Thus we might expect that all eigenstates with
sufficiently small $j$ would have the very same values of the
parameters for large enough $n$ in the ISR. Table 1 serves to check
this point.  In this table we have shown values of the ratio of these
parameters for various different eigenstates. According to the
theory, the ratio should be unity.  Clearly the theory works well for the
magnitude eigenstates. (The table also shows that the constancy of the
parameters does not work for the phase eigenstates, but that story will be told
later.)

Thus, the matching into the VSR uses two of the four boundary conditions on our
linear chain problem.  A very similar mechanism sets the other two boundary
conditions in the regions in which  $n$ is of order $j$.  There are two
quantities which can determine the behavior of the eigenfunction in this
region.  The first is the variation in helicity coefficient $B$ produced by
the disturbance.  This variation produces a term in $\delta \Phi_{n}^j$
which varies as $(\eps - 1)^n$ and which then grows to be of relative
order unity when $n$ is of order $j$.  By setting this coefficient,
and also the value of the $\sigma ^j$, one has two coefficients at ones
command. These two are just enough to ensure that the eigenfunction decays,
rather than grows, as $n$ goes to one.

There is only one important difference between the problem of
determining the static solution $u_n$ and the eigenfunctions.  For the
former, we have forcing on the first shell. In the latter, the
important forcing is
produced by the eigenvalue term, and occurs on the $j^{th}$ shell. The scaling
analysis for the  $j^{th}$ eigenvalue is modified because the range of the
cascade becomes $j$ to $N_D$ rather than the $1$ to $N_D$ that we used in the
analysis of the velocity.  Thus the scaling rule, with corrections, becomes on
the small-$j$ end
\alpheqn
\begineq
{\sigma'}^{(j)}=\sigma^{(j)} (1+ O(\frac{k_j}{k_{N_D}})^{4/3}))
\quad\hbox{for $j$ in region S}
\label{eiS}  \endeq
Here, as before, the prime indicates a decreasing in the viscosity by a factor
of $\lambda^4$ together with a shift in the cutoff, $N$. The
corresponding result on the large-$j$ end is the statement
 \begineq
\sigma'^{(j+3)}=\lambda^2\sigma^{(j)}(1+ O(\eps-1)^j) \quad\hbox{for $j$ in
region L}  \label{eiL}
\endeq
\reseteqn
Both together give the scaling law:
\begineq
{\sigma'}^{(j+3)}=\lambda^2 \sigma^{(j)}
\quad\hbox{for $j$ in region I} \label{eiI}  \endeq
To check our analysis, we should check these rules.  The first check, done in
figure \ref{over1m}a,  is to plot overlays of the primed and unprimed
eigenvalues on polar
plots like that in figure \ref{spec}. 
We show the phase eigenvalues here, but the scaling fits are even better for
the magnitude eigenvalues.
The two spectra overlay precisely at small
$j$, as expected, and fit badly for large $j$.  In the second check, one plots
overlays
of $\sigma$ and $\sigma'/\lambda^2$ as shown in figure \ref{over1m}b.  This
overlay shows
agreement between the two for large eigenvalues but not for small. A more
careful check is to take the ratio of the nearby eigenvalues in these two
figures. Call this ratio $R$.  The quantity $|R|-1$ is a quantitative measure of
the errors in our statements (\ref{eiS}) and (\ref{eiL}). Fig \ref{ew-error}
plots this ratio versus
j and shows that the order errors vary as stated in these equations.

Thus, one can feel that the magnitude response is understood reasonably well.

\subsection{Eigenvalues and Conservation Laws}
One basic principle about this model is that the dissipation can play a large
role in determining the ISR behavior. To see this fact in more detail, we
examine the effect of the conservation laws for energy and helicity upon the
eigenvalue analysis.

Let us recollect that the conservation laws for the the system can be
expressed as
\begineq
\sum_n U_n C_n(U) h^n=0.
\label{eq30}
\endeq
Here $h$ is a quantity which defines the two conservations.  For energy
conservation $h=1$; for helicity
\begineq
h=(\eps - 1)^{-1}.
\label{h}
\endeq
For the
$j^{th}$ right eigenvector $\delta \Phi^j$ and its eigenvalue
$\sigma^j$, we have an eigenvalue equation of the form  \begineq
\sigma^j \delta \Phi_n^j  =
\sum_m (\pm C_{nm} - D_{nm} ) \delta \Phi^j_m.
\label{eigen}
\endeq
Here the minus sign corresponds to the magnitude eigenvalues and the plus
to the phase eigenvalues. Multiply by $(u_n)^2 h^n$ and sum over all $n$ to
obtain
\begineq
\sigma^j \sum_n (u_n)^2 h^n \delta \Phi^j_n =
\pm \sum_{nm} h^n (u_n)^2 C_{nm}(u) \delta \Phi^j_m - \nu \sum_n h^n (u_n)^2 k_n^2\delta
\Phi^j_n. \label{eigen2} \endeq
The  conservation identity (\ref{eq30}) is true for any $U$ and hence also
for $U=u(1+\delta \Phi)$. Since $\delta \Phi$ is a small perturbation
 we can expand around $u$.
\begineq
0=\sum_n h^n U_n C_n(U)=
\sum_n h^n [u_n C_n(u)[1+  \delta \Phi^j_n ]
+ \sum_{nm} (u_n)^2 h^n C_{nm}(u) \delta \Phi^j_m ].
\label{eq31}
\endeq
In this way we have expressed the Jacobian (\ref{eq5b}) in terms of the
cascade itself.
(\ref{eq30},\ref{eq31}, and \ref{eq1}) yield
\begineq
-\sum_{nm} h^n (u_n)^2 C_{nm} \delta  \Phi^j_m =
\sum_n h^n u_n C_n \delta  \Phi^j_n=\sum_n(-D_n+F_n) u_n h^n \delta \Phi^j_n.
\label{eigenid}
\endeq
Now we can rearrange the eigenvalue equation.  In the case of
magnitude response the dissipation term in equation (\ref{eigenid}) adds to the
dissipation term in equation (\ref{eigen}) to give us an identity for the
eigenvalue:
 \alpheqn
\begineq \sigma_M^j\sum_n h^n (u_n)^2 \delta \Phi^j_n=
- 2 \nu \sum h^n (u_n k_n)^2\delta \Phi^j_n + f h u_1 \delta \Phi^j_1.
\label{leo1a}
\endeq

The left hand side of this equation is the rate of decay of the conserved
quantity, as determined by the eigenvalue.  On the right we see that the decay
is (naturally enough) not produced by the cascade but only by the dissipation
through viscous damping and also by the addition through the external force,
$f$.  Thus we understand once more that the dissipation must have a crucial
role to play.

If we go through the same calculation for the phase response, the result is
totally different.  Instead of adding to one another, the dissipation terms
cancel out (!), leaving us with the identity

\begineq \sigma_\phi^j\sum_n h^n (u_n)^2 \delta \Phi^j_n=
- f h u_1 \delta \Phi^j_1.
\label{leo1b}
\endeq
\reseteqn

Compared to the magnitude sector,
the phase sector shows a quite different form for the conservation law
identities. We can no longer say that dissipation produces decay. Instead we
say that the relevant quantity is added through the force term and then
changes in time because of this addition.   We now turn to a more detailed
consideration of the phase sector.

\section{Phase Response}

\subsection{Establishment of phase transitions}

{}From what we have seen, both $u_n$ and the entire magnitude sector of the
linear response vary smoothly as $\eps$ passes through the Biferale value given
by equation (\ref{luca}). When $\lambda = 2$, this transitional value is $0.37$.
The story is different for the phase response. Whenever $\eps$ passes
through the
critical value, then there is a quite apparent change in the structure of the
eigenvalues. As the viscosity goes to zero,  this
change involves having a large number of eigenvalues pass from being
stable to being unstable. In the asymptotic limit,  one third of the entire set
of ISR eigenvalues undergo such a passage at this point.

To see the evidence for this proposition, return to part b of figures
\ref{spec},\ref{spec.5}, and figure \ref{specc}. These pictures respectively apply
below, above, and at the phase transition in $\eps$. Away from the phase
transition, the phase eigenvalues fall into two classes: The
eigenvalues fall onto three lines of constant phase: one real and two with
opposite phases. Others, the ``deviating eigenvalues'' do not seem to
fall into the simple pattern, and have phases which change from eigenvalue to
eigenvalue. Since we are interested in scaling properties, we want to know
something about the pattern which arises when the viscosity is taken to zero. An
examination of the spectrum shows that its three main branches grow longer as
$\nu$ becomes smaller, but that the number of deviating eigenvalues does not
grow. Figure \ref{ntoinf} presents a counting of eigenvalues in the phase
response sector. They are divided into the following categories:  real,
"constant phase" complex, and deviating eigenvalues. They are then counted at
different values of $N_D$. We see that the number of deviating
eigenvalues remains the same as we change the number of shells in the ISR. The
same analysis --and result-- applies above the transition. Consequently, for
$\nu \to 0$ the number of deviant eigenvalues becomes negligible compared to the
number in the three main branches. In the asymptotic limit, each of the
main branches has one third of the total eigenvalues. Thus, the deviating
eigenvalues are not a scaling limit phenomenon, but rather a transient which
defines the approach to scaling at the ends of the ISR.

There is a main branch on the real axis both above and below the transition.
Above the transition, this branch is unstable; below it is stable. At the
critical point, we have a change involving, in the $\nu \to 0$ limit, an
infinite number of eigenvalues.

Figure \ref{scenario}, together with figures \ref{spec}b,\ref{spec.5}b, and
\ref{specc}, presents the flow of phase 
eigenvalues as $\eps$ goes from below to above the transition. We start
at \ref{scenario}a with the familiar spectrum at $\eps=0.2$. 
Real eigenvalues collide and turn to 
complex ones and these (so created) deviating group of eigenvalues wanders
towards the imaginary axis. If we had chosen a larger system the number
of deviating eigenvalues would still be the same, and only the straight 
branches would gain in members.    
The smaller the system is the more is the discontinuous
transition smeared out by finite size effects. (For finite $\nu$ also the
critical point $\eps_c$ weakly depends on $\nu$ and slightly deviates from
$\eps_{bif}$, see figure \ref{epsofnue}. These finite size effects spoil the
regularity of the phase transition. A related transition with less
disturbing finite size effects is discussed in appendix B)

As the deviating group is close to the phase transition the two conjugate
straight branches of eigenvalues split into two (Fig. \ref{specc}), causing a 
breakdown of
scaling law $S$. A scaling of the form $S^2$
(i.e., make use of the symmetry $n\to n+6$) however,
is still valid. The scaling of the
solution and of the magnitude eigenvalues does not break down at this point.

Above the transition fig \ref{scenario}d the deviating group has curved in
the other direction and the eigenvalues return to the now unstable side of
the real axis.

There is a hidden structure of the deviating group not easily seen in our 
representation (\ref{r}); these eigenvalues fall on a straight line in a 
$log(|\sigma|)-arg(\sigma)$
plot. Figure \ref{angle} shows (the upper half) of the deviating group and 
the real eigenvalues for
$\eps=0.33,0.37$, and $0.4$. A dotted line marks the border of instability
at $\pi/2$.
Remember again that only branches with constant phase (here horizontal)
will contribute in
the $N \to \infty$ limit. From figure \ref{angle} it is evident that with
$N \to \infty$ instability will occur immediately above $\eps_{bif}$, since an 
arbitrary small tilting suffices to produce unstable real eigenvalues.
Moreover, for $\nu \to 0$ there is not even a finite number of unstable
eigenvalues for $\eps<\eps_{bif}$ (see Fig. \ref{epsofnue}). 
This means that $\eps_{bif}$ coalesces with $\eps_c$:
\begineq
\eps_{bif}=\eps_c.
\label{eqtrans}
\endeq

Although the instability mechanisms consists of one pair
of complex conjugate eigenvalues after another crossing the imaginary axis,
at $N \to \infty$ an infinite number of such pairs coalesce and an infinite
number of oscillatory instabilities is unleashed within an
arbitrarily small change
in $\eps$.

For $\nu \to 0$ the destabilization scenario of the fixed point is thus
different than has been suggested in the light of "finite-size simulations"
(\cite{bif95,sch95}).
On one side of the phase transition we have totally stable behavior, while
on the other side we have immediately an infinite number of unstable modes
on all scales. These instabilities are purely exponential and not oscillatory.

\subsection{Asymptotics}

The phase transition is a change in the behavior of a very large number of
eigenfunctions all at once. Biferale \cite{bif95b} has explained this phase
transition as a blockage in the energy flow caused by a flow of helicity.
Since the blockage can occur anywhere in the ISR it seems reasonable to
assume that, when the conditions are right, blockages can occur at many
places affecting many eigenfunctions and eigenvalues at once.

Another way of asking the same question is more mathematical in character.
To see this phase transition we must have some kind of change in the
small-$\nu$ asymptotics of the system.  That is we must have some sort of
change which can be seen by an eigenfunction with $j$ much larger than unity
and much smaller than $N_D$. So we need an asymptotic theory of eigenfunctions
for this system.

To understand the phase transition, we must first understand how the phase
sector can be so unstable. The key is given in Table 1. In that table we
see that in contrast to the magnitude sector, the phase sector shows far more
flexibility in the large-$n$ limit of the eigenfunction. The magnitude sector
has but one value of $p_1$ and $p_2$ for all eigenfunctions in this region.
The phase sector can actually have two different linear combinations. The two
permitted combinations are:
\begineq
\delta\Phi_{3k+1}=0   \quad  \delta \Phi_{3k+2}=-\delta \Phi_{3k}
\label{exact}
\endeq
and also, for example,
\begineq
-\frac{1}{2} \delta \Phi_{3k+1}=\delta \Phi_{3k+2}=\delta \Phi_{3k}
\label{approx}
\endeq
for integer values of $k$. The changes in expression (\ref{exact}) are an exact
symmetry of the GOY system and produce a phase sector eigenfunction which has
eigenvalue zero. The changes in expression (\ref{approx}) satisfy the eigenvalue
equation for all $n$-values except $n=1$.  Here, the eigenvalue equation fails
because of the forcing term. However, this combination also forms a possible
high-$n$ behavior. All eigen functions have one or another linear
combination of these two behaviors as the high-$n$ limit. But there are two
such linear combinations possible.  In comparison to the magnitude case, we seem
to have lost a boundary condition at the high-$n$ end.

Thus, we state the first
contrast between the magnitude sector and the phase sector. The magnitude
sector can be described by giving two boundary conditions at the high-$n$ side
of the ISR; the phase sector can be described by giving only one.

We must have someplace an extra boundary condition for the phase sector.
Bifarale directed our attention to the conservation of helicity.  Instead
of doing a full analysis of the possible extra conditions, we just follow
his direction and look at the helicity conservation law, equation
(\ref{leo1b}), in
the phase sector. (The reader will recognize that some leap of faith is required
around this point in the argument.)  This equation is
 \begineq
\sum_n h^n (u_n)^2 \delta \Phi^j_n=
 - {f h u_1 \delta \Phi^j_1 \over \sigma_\phi^j}  \nonumber.
\endeq
But, for an eigenvalue with $j$ in the middle of the ISR, the first component
of the eigenfunction is very, very small. This component goes to zero more
rapidly than an exponential in $j$.  For this reason, it is quite
reasonable to neglect the right hand side of the helicity identity and
write instead:
\begineq
\sum_n h^n (u_n)^2 \delta \Phi^j_n= 0 \quad \mbox{for $j$ in region I}
\label{helid}
\endeq
We propose that this identity replaces the lost high-$n$ boundary condition 
for the asymptotics of the phase sector.

This proposition has several quite attractive features. We expect the
eigenfunction to be of roughly the same  order of magnitude over the
whole range in which $n$ is greater than $j$.  Then for
$\eps<\eps_{bif}$ the summation has its main contribution at the lower-$n$
end and then falls off as
\alpheqn
\begineq
\delta \Phi^j_n \left(-{1-\eps_{bif} \over 1- \eps}\right)^n.
\label{low}
\endeq
When we are at the critical point, the summation does not fall at all
but the summand looks like
\begineq
\delta \Phi^j_n (-1)^n.
\label{critp}
\endeq
\reseteqn
Finally, above the critical point the major contributions come at the
high-$n$ end and then fall off toward lower $n$ with the same law as shown in
equation (\ref{low}).  
Near the phase transition, these behaviors produce long-range
correlations between the different parts of the ISR.  It is these
correlations which form the key to the phase transition.

{}From these assumptions, one can find scaling laws for the phase
eigenvalues, namely on the small-$j$ end
\alpheqn
\begineq
{\sigma'}^j=\sigma^j \left(1+ O\left(
{{1-\eps_{bif}} \over {1-\eps}}\right)^{N_D-j}\right)
\quad\hbox{for $j$ in region S}
\label{eiSp}  \endeq
Here, as before, the prime indicates a decrease in the viscosity by a factor
of $\lambda^4$ together with a shift in the cutoff, $N$. The
corresponding result on the large-$j$ end is the statement, which we have used
before.
\begineq
\sigma'^{(j+3)}=\lambda^2\sigma^{(j)}(1+ O({\eps-1)^j})
\quad\hbox{for $j$ in region L}  \label{eiLp}
\endeq
\reseteqn
Both together give the scaling law:
\begineq
{\sigma'}^{(j+3)}=\lambda^2 \sigma^{(j)}
\quad\hbox{for $j$ in region I} \label{eiIp}  \endeq

To check the accuracy of our thinking we show in figure \ref{over1p}
polar plots in which we
overlay the eigenvalues $\sigma_j$ with, respectively, the eigenvalues
$\sigma_j'$ and
$\lambda^{-2}\sigma_{j+3}'$.  According to the theory, the first of these
should agree very well
for small magnitudes of the eigenvalue, while the second should agree on
the large
magnitude end. The figures bear this out. For a more accurate check, we
construct errors as,  for example,
\begineq
\delta = \left| {\sigma_j \over   \sigma_j'}\right| -1.
\endeq
These errors are shown in figure \ref{ew-error} and compared with the
theoretical estimates taken from equations (\ref{eiLp}) and (\ref{eiSp}).
Since the estimated and the actual errors agree rather well, we can
argue that we have caught the essence of the phase eigenvalues.

\section{Summary and Conclusions}
The GOY model has a linear response behavior about a static solution in
which the
stability eigenvalues extend over a huge range of frequencies. In the ISR,
the eigenvalues show a scaling behavior limited by disturbances from stirring
or from viscous effects.  The scaling of the eigenvalues is more complex in
the phase sector than in the magnitude response.  The phase sector shows a
phase transition connected with a boundary condition which moves from one
end of the ISR to the other. This boundary condition is derived from the
conservation law for the model's version of helicity.  Eigenvalues change quite
abruptly as the model's parameter passes through its phase-transition value.

Much of the
scaling mirrors the scaling properties of the static GOY model.  However,
there is scaling associated with the phase transition which we have not
investigated in any detail.  We have also not fully established the scaling
behavior of the linear-stability eigenvectors. These studies are left for the
future.

For now, we have demonstrated scaling in the linear response of a still system.
We have seen a new phase transition and gained a qualitative understanding of
its source.

\vspace{1.5cm}
\noindent
{\bf Acknowledgements:}\\
It is our pleasure to thank Luca Biferale,
Jean-Philippe Brunet, Peter Constantin, Greg Huber, and Norman Lebovitz
for helpful discussion. We are also indebted to Peko Hosoi, who programmed
our IDL animations.
This work has been supported by DOE contract number DE-FG02-92ER25119,
by the MRSEC Program of the National
Science Foundation under Award Number DMR-9400379, and
by the Deutsche Forschungsgemeinschaft (DFG) through its SFB185.

\newpage

\appendix

\section{Large scale forcing}

Most studies of the GOY model
\cite{jen91,kad95,sch95} employed $F_n= f \delta_{n,4}$
as large scale forcing.
Though this forcing seems innocent,
it has some disadvantages. The
velocity component $u_3$ is much smaller than $u_{1,2,4,5,6,\dots }$
\cite{sch95}, what is clearly unphysical.
We force the system on level $n=1$. As can
be seen from equations (\ref{eq1}) and (\ref{eq2}), with
$u_3=0$ and $n \to n-3$,
the shift in the forcing by three shells
just shifts the solution by three shells as well.
{\footnote{ 
With the forcing $F_n \propto \delta_{n,1}$ a static solution of eq.
(\ref{eq1}) exists {\it for all} $0<\eps<1$,
i.e., the saddle node bifurcation found
in \cite{sch95} is an artifact of the forcing
 $F_n\propto \delta_{n,4}$.
In particular,
the overall existence of the static solution now allows for an
analysis of the eigenvalue spectrum for the parameter values
$\lambda=2$ and $\eps=0.5$
\cite{yam87,jen91,ben93}.
}} 
{\footnote{  
The artificially small $u_3$ (for $F_n\propto \delta_{n,4}$-forcing)
also leads to strongly non-normal eigenvectors. This was
first discovered by J.P. Brunet \cite{bru95},
who also calculated the corresponding pseudoresonance.
The reason for that can be seen
from (\ref{eq5b},\ref{eq4b}): $(A_\phi )_{3,m}$ with $m=1,2,4,5$ is
small as $u_3 \sim \nu$ is so small.
All the eigenvectors of $A_\phi$ then have a
huge $3^{rd}$ component and are thus nearly parallel to each other.
This makes $A_\phi$ very
nonnormal, the more, the smaller $\nu$ is.}} 
 
With the different forcing the borderline of stability becomes only
slightly different. For instance, $\eps_c$ beyond which
(\ref{eq1}) becomes unstable is in the standard case
($\lambda=2, \nu=10^{-7}, f=\sqrt{2} \cdot 5 \cdot 10^{-3}$)
shifted from 0.370 to 0.377.
More significant is the change
in the amplitude of the $\eps_c$ oscillations with viscosity.
This viscosity dependence
is shown in Fig. \ref{epsofnue} for $\lambda=2$, and looks similar
for other values of $\lambda$.
In contrast to the case with
forcing on the 4th shell (see \cite{sch95}, Fig. 12) the
stability border $\eps_c$ now seems to approach
the constant $\eps_{bif}$ for vanishing viscosity $\nu\to 0$,
\begineq
\lim_{\nu \to 0}{\eps_c}=\eps_{bif}:= 1-\lambda^{-2/3}.
\endeq
Some of the transitions to instability in Fig. \ref{epsofnue} are due to a 
complex pair of eigenvalues passing
through the imaginary axis, others are due to a single real eigenvalue
going through zero.
{\footnote{
The model with the forcing on the fourth shell shows a transition from
stable to unstable behavior only via two complex conjugated eigenvalues
going through the imaginary axis. With the forcing
on the first shell instability for some $\nu$ occurs via a real eigenvalue
going through zero, followed by a series of Hopf bifurcations
(take e.g. $\nu=10^{-8}, f=\sqrt 2 \cdot 0.005, k_0=0.0625$).
However, from the viewpoint of the
scenario of instability in the limit $\nu \to 0$,
treated in section 4.1, this additional type
of transition is not fundamentally different from the other one.}}

In the main text we have studied the behavior of the overall eigenvalue
spectrum. The stability border, on the other hand, is determined only by
the first eigenvalue crossing the imaginary axis. Nevertheless,
in the main text we have two strong evidences for a dependence of the
stability border on the particular forcing. 
As we have seen in section 3.2 the modulus of the eigenvalues 
is set by the cascade term.
The eigenvalues that decide the stability are the small ones close
to zero. Their size should depend on how the solution looks like at the
very first shells, and this is obviously influenced by the kind of forcing.
Also formula (\ref{leo1b}) shows that the stability border should
depend on the particular forcing.

\section{A related phase transition}
In this appendix we analyze the related phase transition in the spectrum to the
matrix
\begineq
A_\xi = \xi \cdot C -D .
\label{pertur}
\endeq
The matrices
$C$ and $D$ contain the solution of (\ref{eq1}). For general $\xi$ the matrix
$A_\xi$ is {\it not} the stability matrix of the system.
Only 
for $\xi =-1 $ we have $A_{-1} = A_M$ and for $\xi=1$ we have
$A_{1} = A_\phi$.
The generalized amplitude matrix $A_\xi$ with $\xi <0$ shows the same
features as $A_M$ itself. So we only  study the interesting case
of  $A_\xi$ with
$\xi >0$ which displays a {\it phase transition} at $\xi =1$.
The advantage of studying this phase transition rather than the one in $\eps$
discussed in the body of the paper is that here the critical point $\xi=1$ does
not depend on the viscosity. Thus finite size effects are less
complicated than
for the transition in the spectrum of $A_\phi$
at the $\nu$ dependent $\eps_c$.
Taking $\xi\approx 1$ (but $\xi\ne  1$)
can be thought of as equivalent to slightly varying the
dynamical equation (\ref{eq1}) and thus changing the phase symmetry of
that equation, which leads to a different solution.
We perform the analysis for $\eps=0.3$ as it is the spurious
stabilization of the phase matrix for $\eps<\eps_c \approx 0.37$ which we want
to understand.

We start with $\xi=0$. All eigenvalues $-\nu k_n^2$ of $A_0=-D$ are of course
stable. However,
with increasing $\xi$,
more and more
eigenvalues of $A_\xi$ turn {\it positive} as soon as $\xi \ageq \nu$.
This feature holds for $|\lg \nu | \sim 30$ orders of magnitude in $\xi$.
{}From comparing the viscous and the inertial contribution to $A_\xi$ we obtain
that the largest eigenvalue increases as
\begineq
\sigma_{max} \sim
\xi^{3/2} \nu^{-1/2}.
\label{estim2}
\endeq
$\sigma_{max}$ becomes as large as $10^{12}$ for $\xi\approx 0.63$ and the
standard parameters $\lambda = 2$, $\eps = 0.3$, $N=81$, $\nu \sim 10^{-30}$.
This is the behavior we see in the spectrum of $A_\phi = A_1$ for $\eps >
\eps_c$ and what we had expected also for smaller $\eps$.

Why and how is stability achieved for $\xi=1$, i.e., why does
$A_{1} = A_\phi$ not have any positive eigenvalue
for $\eps < \eps_c$?

For $\xi$ growing beyond $0.63$ towards 1,
two real eigenvalues merge, form a complex pair which moves in
the complex plane on a circle and finally turns stable through an inverse Hopf
bifurcation. This happens again and again through a selfsimilar cascade of
bifurcations towards phase symmetry at $\xi = 1$ -- no unstable eigenvalue is
left. This phase symmetry of $A_1 = A_\phi$, discovered in \cite{sch95},
reflects itself in a zero eigenvalue, as discussed above.

We  now analyze the singularity in detail. We introduce the
distance $\tau$ from the singularity at $\xi=1$,
\begineq
\xi=1 +\tau
\label{tau}
\endeq
and increase $|\tau|$ on a log scale from $\tau=0$. In this way we
break the phase symmetry in a stronger and stronger way.

We describe the features for $\tau \le 0$.
For $\tau=0$ we only have negative eigenvalues and one eigenvalue
equals zero. This center manifold eigenvalue turns {\it positive} for
$|\tau| > 0$, signaling instability. Moreover, two different small
(modulus wise) real  eigenvalues ($<0$) merge on the real axis and
form a complex pair which wanders towards the $\Re\sigma = 0$ axis and
turns unstable via a Hopf bifurcation. This happens as early as for
$\tau\approx 10^{-10}$, see figure \ref{aehre}.
In the right complex half plane they continue to wander on a kind of circle,
finally merging again on the real axis, now forming two positive real
eigenvalues. But meanwhile a second pair of complex eigenvalues has
formed in the left half plane which again turns unstable via a Hopf
bifurcation and then finally merges. This happens again and again,
leading to $3,5,7,9,11,\dots $ positive eigenvalues.
The real parts of the positive eigenvalues are plotted in fig.\ \ref{aehre}.
The selfsimilarity of the cascade of bifurcations can clearly be observed. The
maximal eigenvalue $\sigma_{max} \propto \tau^{4/3}$. The scaling law mirrors
the classical K41 scaling of $u_n$.

For $\tau > 0$
the behavior is very similar to the $\tau < 0$ case.
The only difference is that the center manifold
(zero) eigenvalue turns stable first
and we now have an {\it even} number  
$2,4,6,8,\dots$ of positive eigenvalues.

To demonstrate the finite size effects of the transition we plot
the number of positive eigenvalues as a function of $\lg\tau$ for various
system sizes $N$, i.e., viscosities $\nu$, see figure \ref{finitesize}.
In the  $N\to \infty$ limit, even for an arbitrarily small $|\tau |$,
we have  an infinite number of unstable eigenvalues.

\newpage

\centerline{\bf Table}

\begin{table}[htp]
\begin{center}  
\begin{tabular}{|c|c|c|c|c|c|}
\hline   
$j$& $n$&\multicolumn{2}{|c|}{magnitude}&\multicolumn{2}{|c|}{phase}\\
\hline
{}&{}&$Re({p_1^j(n)\over p_1^{j+1}(n)})$&$Re({p_2^j(n) \over p_2^{j+1}(n)})$&
      $Re({p_1^j(n)\over p_1^{j+1}(n)})$&$Re({p_2^j(n) \over p_2^{j+1}(n)})$\\
\hline
18 & 35 &  1.02897 &  0.99597 & 0.02867  &  0.81927 \\
18 & 45 &  1.00132 &  1.00041 & -4.83889 &  22.2840 \\
18 & 50 &  0.99986 &  1.00002 & 0.02867  &  0.81891 \\
18 & 55 &  1.00004 &  1.00002 & 0.31876  & -0.00242 \\
18 & 62 &  1.00000 &  1.00000 & 0.02867  &  0.81891 \\
18 & 70 &  1.00000 &  1.00000 & 0.31876  & -0.00242 \\
\hline
13 & 16 & 75.5759 & 2.08638 & -0.58804 & -51.81228 \\
13 & 35 & 0.99648 & 1.00048 & 40.53829 &  -0.11043 \\
13 & 45 & 0.99984 & 0.99995 & -0.00677 &   0.02073 \\
13 & 55 & 1.00000 & 1.00000 & -0.42772 & -51.26321 \\
13 & 71 & 1.00000 & 1.00000 & 40.53836 &  -0.11044 \\
\hline
\end{tabular}
\end{center}
\caption[]{
Behavior of parameters for eigenstates. 
Here $j$ refers to the $j^{th}$ eigenfunction while $n$ describes the
components of that eigenfunction.
These two parameters describe the
large-$n$ behavior in the ISR. They are fixed by the value of 
$\nu=10^{-3}16^{-21}$ in the
magnitude sector but vary considerably in the phase sector. 
The ratios are essentially the same for the n-values
in between the given values.
(Correlations
between phase values with $n$ different by a multiple of 3 stem from the period 3
in the eigenvectors.)
} \label{table1}
\end{table}
\vspace{0.5cm}

\newpage
\centerline{\bf Figures}

\begin{figure}[htb]
\caption[]{
Linear Stability Eigenvalues. The case considered is $\eps=0.3, \nu=10^{-3}16^{-26}, N=90$.
Here we plot the amplitude (diamonds) and phase (circles)
of the eigenvalues in a kind of polar plot. The phase in this polar diagram
IS the phase of the eigenvalue.  The radial coordinate is given by equation
(\ref{r}) which essentially produces the logarithm of the eigenvalue.
} \label{spec}
\end{figure}
\vspace{0.5cm}

\begin{figure}[htb]
\caption[]{
The same as Figure \ref{spec} except that now $\eps$ has the value 0.5.
The amplitude matrix becomes eventually unstable at $\eps \approx 0.558$.
}\label{spec.5}
\end{figure}
\vspace{0.5cm}

\begin{figure}[htb]
\caption[]{
The same as Figure \ref{spec}b except that now $\eps$ has its critical value,
$\eps_{bif} $ which is about $ 0.37$. Notice that there are now six branches,
all with complex eigenvalues. The equal spacing on each branch is evidence
of scaling.
But, evidently, the scaling is quite different from that shown in figures
\ref{spec}b
and \ref{spec.5}b
} \label{specc}
\end{figure}
\vspace{0.5cm}

\begin{figure}[htb]
\caption[]{
We plot
$W_n=k_n^{1/3} U_n$
against $n$. Notice how $W_n$ oscillates in the inertial range and how it
falls off quite rapidly in the dissipative range. (The latter
is for $n > N_D \approx 67$.) The actual calculation is done for
$\eps=0.3, \nu=10^{-3} 16^{-20}$ (circles) and $\nu=10^{-3} 16^{-21}$ (dots).
The behavior of W in these two simulations is compared. 
In the second (dots), N is increased by three and $\nu$ is decreased
by a factor of $\lambda^{-4}$. For all $n$ smaller than 56 or
so the two calculations agree.
(b) A comparison like that in a) except that now $W_n$ is
compared with $W'_{n-3}$. These agree top plotting accuracy for
all $n$ bigger than 20 or thereabouts.
} \label{sol}
\end{figure}
\vspace{0.5cm}

\begin{figure}[htb]
\caption[]{
The error in the scaling relations for the velocity. The deviation
from unity of the ratio of two pieces of data in equations (\ref{delus})
(triangles) and (\ref{delul}) (diamonds). The slope of the solid lines show
the theoretical estimates of the error.
} \label{solerr}
\end{figure}
\vspace{0.5cm}

\newpage

\begin{figure}[htb]
\caption[]{
The general structure of right eigenvectors for the
the amplitude $\delta \Phi_n^{(26)} u_n k_n^{1/3}$ (solid) and
phase $\delta \Phi_n^{(26)}$ (dashed line)
($N=75, \eps=0.3, \nu=10^{-3}16^{-21}$).
In both cases the modulus of the eigenvector is displayed.
The amplitude eigenvector gets damped when it enters the viscous range,
while the period 3 of the phase eigenvector continues into the viscous range.
}
\label{someevs}
\end{figure}
\vspace{0.5cm}

%\begin{figure}[htb]
%\caption[]{
% Eigenvalues in the magnitude sector for $\eps=0.3, N=90$, and $N=93$ using
%the same
%'polar'  representation as in figure \ref{spec}. These eigenvalues are
%overlaid with
%'primed' eigenvalues with the same value of $j$. b) The same data, but the
%primed eigenvalues 
%are multiplied by a factor of $\lambda^{-2}$.
%} \label{over1m}
%\end{figure}
%\vspace{0.5cm}

\begin{figure}[htb]
\caption[]{
a) Eigenvalues of the phase matrix for $\eps=0.3, N=90$, and $N=93$.
One can see
three main branches and the validity of scaling law S.
b) The same data, but the eigenvalues for $N=93$ are shifted by 2.
Scaling law L is at work.
} \label{over1p} \label{over1m}
\end{figure}
\vspace{0.5cm}

\begin{figure}[htb]
\caption[]{
Errors in the scaling laws for the eigenvalues, equations (\ref{eiS},\ref{eiSp})
and (\ref{eiL},\ref{eiLp}). Here we plot
$|R|-1$  where $R$ is the ratio of the two sides of the equation.  The
diamonds describe the
ratio in equations (\ref{eiS},\ref{eiSp}) while the triangles do the same for
equation (\ref{eiL},\ref{eiLp}).
The unfilled symbols are for the amplitude matrix and the filled ones stand
for the phase matrix.
Lines are theoretical estimates of the errors.
Part (a) is for $\eps=0.3$ and part (b) for $\eps=0.5$.
The errors are for the modulus of the eigenvalues, but the
errors in $arg(R)$ follow the same trends.
} \label{ew-error}
\end{figure}
\vspace{0.5cm}

\begin{figure}[htb]
\caption[]{
Number of real (solid line), "straight" complex (dashed), and deviating
eigenvalues (dots) with increasing shell number $N_D$.
The slope of the lines is exactly $2/3, 1/3$ and $0$ respectively.
}
\label{ntoinf}
\end{figure}
\vspace{0.5cm}
 
\newpage

\begin{figure}[htb]
\caption[]{
This sequence of pictures shows the phase eigenvalues with changing $\eps$.
We start out with three completely stable branches in (a). Real eigenvalues
turn complex and move towards the imaginary axis (b) until fewer and fewer real
eigenvalues are left (c). Notice, all the 'deviating' eigenvalues that do not 
lie on a straight line are finite size effects, i.e. they constitute only a 
negligible part for a sufficiently large matrix. These deviating eigenvalues
form a straight line at $\eps_{bif}=0.37$ (Fig. \ref{specc}). 
Now the spectrum consists of
six complex branches and the scaling law (S) is no longer valid and has to be
replaced by scaling law ($S^2$). Above $\eps_{bif}$ (d) eigenvalues return to the
real axis, but are now unstable. Essentially there are two complex branches and
one real branch again as it was below $\eps_{bif}$. An even higher value of
$\eps=0.5$ is shown in figure \ref{spec.5}b.    
} \label{scenario}
\end{figure}
\vspace{0.5cm}

\begin{figure}[htb]
\caption[]{
Main branch of 'deviating' eigenvalues and real eigenvalues
in a $log_{\lambda}|\sigma|-arg(\sigma)$  plot, for
$\eps=0.33$ (triangles), $\eps=0.37$ (squares), and $\eps=0.4$ (circles).
The tilted branch rotates with changing $\eps$ and is horizontal 
at the phase transition and unstable above (black).
It can be seen that in the $N \to \infty$ limit
an infinite number of eigenvalues turns unstable
immediately above the phase transition.
} \label{angle}
\end{figure}
\vspace{0.5cm}

\begin{figure}[htb]
\caption[]{
Border of stability
$\eps_c$ as a function of $\nu$ for $\lambda=2$.
$\eps_c$ approaches the
Biferale prediction (thick horizontal line) for small $\nu$. 
Some of the points are due to a complex pair of eigenvalues passing
through the imaginary axis, others are due to a single real eigenvalue
going through zero.
} \label{epsofnue}
\end{figure}
\vspace{0.5cm}

\begin{figure}[htb]
\caption[]{
All positive eigenvalues of $A_\xi$ for $\tau= \xi-1$ between
0 and $-1$ on a log-log scale in $|\tau|$, corresponding to a
lg vs $\lg\lg\xi$ plot.
The selfsimilar cascade of bifurcations towards phase symmetry
has its origin in the self similarity of the matrix itself.  
The parameters are $N=80$, $\nu \sim 2 \cdot 10^{-29}$, $\eps= 0.3$,
$\lambda = 2$. For $\tau = -1$ (right edge of the plot)
we have $A_\xi = -D$ and all eigenvalues
are again stable.
}
\label{aehre}
\end{figure}
\vspace{0.5cm}

\begin{figure}[htb]
\caption[]{
Number of positive eigenvalues of $A_{1-\tau}$
as a function of $|\tau |$ for various system sizes
$N=80,68,56,44,32,20$, left to right, corresponding to
viscosities $\nu = \nu_0 \cdot 16^i$, $i=0,4,8,12,16,20$,
respectively. $\nu_0 \sim 2 \cdot 10^{-29}$, $\eps = 0.3$,
as in figure \ref{aehre}.
}
\label{finitesize}
\end{figure}
\vspace{0.5cm}

\newpage

\noindent
$*$e-mail: lohse@cs.uchicago.edu

%\bibliography{literatur}

\end{document}